# Magnetic Charge Propagation upon a 3D Artificial Spin-ice

A. May[1], M. Saccone[2], A. van den Berg[1], J. Askey[1], M. Hunt[1] and S. Ladak[*1]

1. School of Physics and Astronomy, Cardiff University, Cardiff CF24 3AA

Email: LadakS@cardiff.ac.uk

2. Physics Department, University of California, Santa Cruz, 1156 High Street, Santa Cruz, CA 95064, USA.

**Magnetic charge propagation in spin-ice materials have yielded a paradigm-shift in science, allowing the symmetry between electricity and magnetism to be studied. Recent work is now suggesting the spin-ice surface may be important in mediating the ordering and associated phase space in such materials. Here we detail a 3D artificial spin-ice, which captures the exact geometry of bulk systems, allowing magnetic charge dynamics to be directly visualized upon the surface. Using magnetic force microscopy, we observe vastly different magnetic charge dynamics along two principal directions. For a field applied along the surface termination, local energetics force magnetic charges to nucleate over a larger characteristic distance, reducing their magnetic Coulomb interaction and producing uncorrelated monopoles. In contrast, applying a field transverse to the surface termination yields highly correlated monopole-antimonopole pairs. Detailed simulations suggest it is the difference in effective chemical potentials that yields the striking differences in monopole transport.**

# Introduction

The concept of magnetic monopole transport within a condensed matter setting has captivated scientists, allowing established theory [1] to become an experimental realization [2-4] within the bulk pyrochlore systems known as spin-ice [5]. In these three-dimensional (3D) systems, rare earth spins are located upon corner-sharing tetrahedra, and energy minimisation yields a local ordering principle known as the ice-rule, where two spins point into the centre of a tetrahedron and two spins point out. Representing each spin as a dimer, consisting of two equal and opposite magnetic charges ($\pm q$), is a powerful means to understand the physics of spin-ice [5]. Using this description, known as the dumbbell model [1], the ice-rule is a result of charge minimisation, yielding a net magnetic charge of zero in the tetrahedra centre ($Q = \sum_i q_i = 0$). Then the simplest excitation within the manifold produces a pair of magnetic charges ($\sum_i q_i = \pm 2q$) which, once created, can propagate thermally and only at an energy cost equivalent to a magnetic analogue of Coulomb's law. The energy scale for the production of monopoles upon the spin-ice lattice is controlled by the chemical potential ($\mu$), which is governed by properties intrinsic to the material such as lattice constant and magnetic moment [6]. Canonical spin-ice materials have a chemical potential that places them in a weakly correlated regime where only a small fraction of bound monopole-antimonopole pairs are found. Recent theoretical work has studied the ordering of magnetic charges upon cleaved spin-ice surfaces, perpendicular to the [001] direction [7]. In such systems, the orphan bonds upon the surface are found to order in either a magnetic charge crystal or magnetic charge vacuum, depending upon the scales of exchange and dipolar energies [7]. Recent experimental work is now hinting at the presence of a surface-driven phase transition [8] but the transport of magnetic charge across such surfaces has not been considered previously.

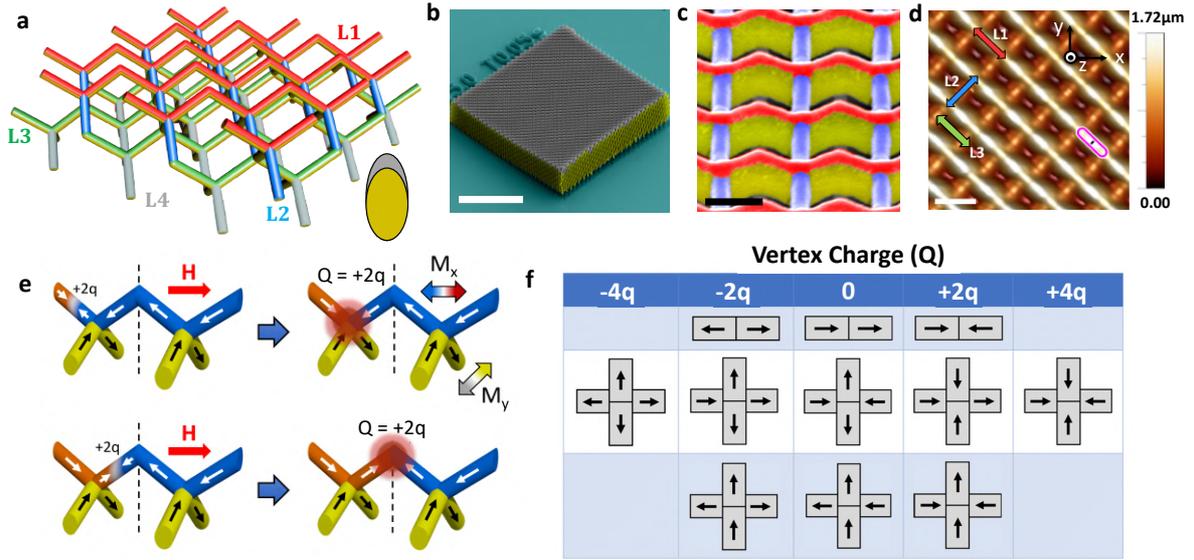

**Figure 1: A 3D Artificial Spin-ice.** (**a**) Schematic of a 3D artificial spin-ice system. The surface L1 layer (red) consists of an alternating sequence of coordination two and coordination four vertices. Below this, the L2 (blue) and L3 (green) layers can be seen. Within these layers, only vertices of coordination four are present. The L4 layer (grey) is the lower surface termination which again has an alternating sequence of coordination two and coordination four vertices. Inset: Cross-section of $Ni_{81}Fe_{19}$ (grey) upon the polymer scaffold (yellow). (**b**) A false colour scanning electron microscopy image of the 3D artificial spin-ice lattice. Scale bar is 20μm. (**c**) A false colour scanning electron microscopy image showing the L1 (red) and L2 (blue) sub-lattices, viewed at a 45° tilt with respect to the substrate plane. Scale bar is 1μm. (**d**) Atomic force microscopy image of the 3D artificial spin-ice system. Scale bar is 2μm. Coordinate system for field application is shown in top-right of image. (**e**) Possibilities for creating magnetic charge upon L1. (**f**) The possible states and associated magnetic charge that can be realised at vertices of coordination two and coordination four.

The arrangement of magnetic nanowires into two-dimensional lattices has recently shown to be a powerful means to explore the physics of frustration and associated emergent physics. These artificial spin-ice (ASI) systems [9-15], where each magnetic nanowire behaves as an effective Ising spin, have recently yielded an experimental realisation of the square ice model [16] and have also been used to study the thermal dynamics of monopoles in the context of Debye-Hückel theory [17]. Controlled formation of magnetic charge is an exotic means to realise advanced multistate memory devices. Such concepts have been shown in simple 2D lattices using magnetic force microscopy (MFM) [18]. The extension of artificial spin-ice into true 3D lattices that capture the exact underlying geometry of bulk systems is paradigm-shifting, allowing the exploration of ground state ordering and magnetic charge formation in the bulk as well as upon the surface. The production of 3DASI systems harbouring magnetic charge also allows marriage with advanced racetrack device concepts [19,20].

In this study, we use state-of-the-art 3D nanofabrication and processing in order to realise a 3DASI in a diamond-bond 3D lattice geometry, producing an artificial experimental analogue of the originally conceived dumbbell model [1]. MFM is then harnessed to image the formation and propagation of magnetic charge upon the 3D nanowire lattice.

**Results**

Figure 1a shows a schematic of the 3DASI, which is composed of four distinct layers, labelled by colour. The system is fabricated by using two-photon lithography [21-24] to define a polymer lattice in a diamond-bond geometry, upon which 50nm $Ni_{81}Fe_{19}$ is evaporated (See methods for further details). This yields NiFe nanowires within a diamond lattice geometry as shown previously [24]. Each nanowire has a crescent shaped cross-section (Fig 1a inset), is single domain and exhibits Ising-like behaviour [24]. The L1 layer, which is coloured red in Fig 1a, is the upper surface termination and consists of an alternating sequence of coordination two (bipods) and coordination four vertices (tetrapods). The L2 and L3 layers, coloured blue and green respectively are ice-like with only vertices of coordination four. Finally, the L4 layer (Grey) is the lower surface termination of the lattice and consists of vertices which alternate between coordination two and coordination four.

The overall array size is approximately 50 μm x 50 μm x 10 μm as seen in the scanning electron microscopy (SEM) image (Fig. 1b). Analysis of SEM data indicates the long-axis of L1 wires is orientated at $\theta = (33.11 \pm 2.94)°$ from the substrate plane, matching within error the angle of 35.25° which is expected for an idealised diamond-bond geometry [5]. A higher magnification image, clearly showing the L1 (red) and L2 layers (blue) can be found in Fig 1c. The topography of the upper three layers can be measured using atomic force microscopy (AFM)

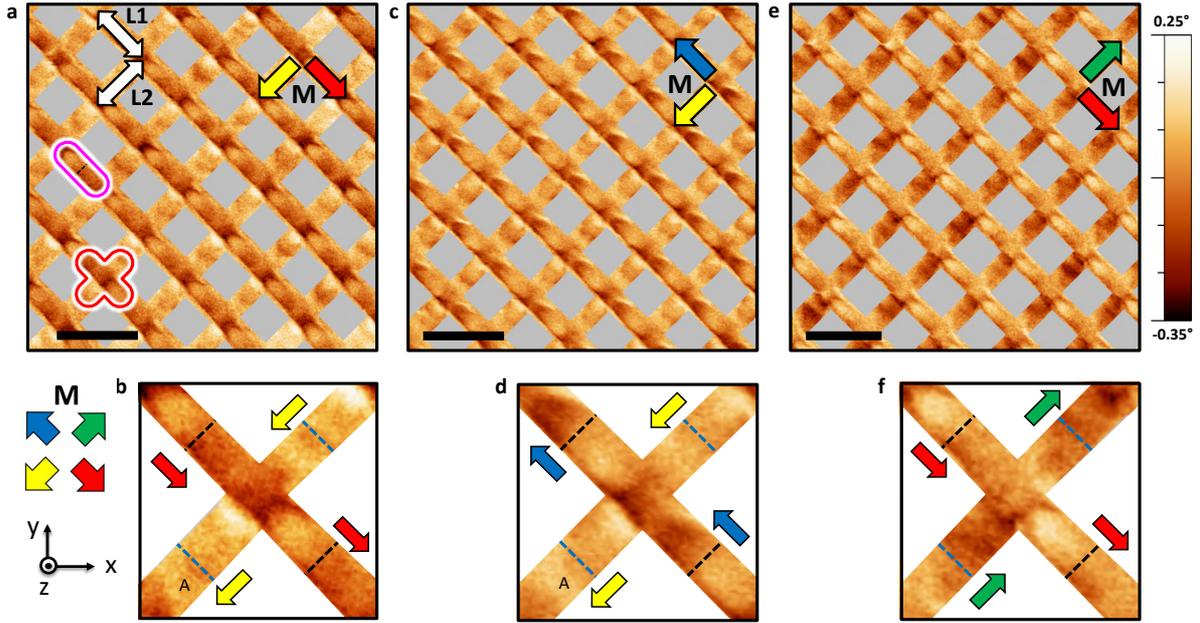

**Figure 2: Imaging the saturated states in a 3DASI.** (**a**) An MFM image taken at remanence after application of saturating fields along the unit vectors (1,-1,0) and (-1,-1,0). A coordination two, surface vertex is highlighted in pink, whilst a coordination four vertex at the intersection of L1 and L2 is highlighted in red. The scale bar represents 2 μm. (**b**) Magnified example of the MFM contrast seen associated with L1-L2 junctions as seen in (a). Arrows are coloured by the local in-plane magnetization components. (**c**) MFM image taken at remanence after a further saturating field is now applied along unit vector (-1,1,0). (**d**) Magnified example of the MFM contrast seen associated with L1-L2 junctions as seen in (c). (**e**) MFM image taken at remanence after a further saturating field is now applied along unit vectors (1,-1,0) and (1,1,0). (**f**) Magnified example of the MFM contrast seen associated with L1-L2 junctions as seen in (e.

as shown in Fig 1d. The coordinate system used to define field directions is also shown in Fig 1d.

The surface of the 3DASI lattice provides interesting possibilities with respect to magnetic charge creation and transport. In Fig 1e we illustrate how magnetic charge propagates along the L1 layer. Starting with a saturated state, applying a magnetic field above a critical value along the unit vector (1,-1,0) leads to the nucleation of a domain wall (DW) (Fig 1e top-left) which carries a mobile magnetic charge of magnitude 2q. When reaching the L1-L2 junction (Fig 1e top-right), the effective vertex magnetic charge becomes Q=+2q. A further increment in magnetic field leads to the L1-L2 junction emitting another DW (Fig 1e, bottom-left) and when this wire is fully switched a surface magnetic charge state of Q=+2q is realized (Fig 1e, bottom-right). Note that a field applied in either direction with a projection along the L2 sub-lattice produces only magnetic charges at four-way junctions (Fig S1). Overall, the 3DASI surface can realise effective magnetic charge of ±4q, ±2q and 0 as summarised in Fig 1f.

**Imaging the magnetic configuration of a 3DASI**

MFM is a convenient method to deduce the magnetization configuration of the 3DASI during field-driven experiments. This imaging technique is sensitive to the second derivative of the stray field with respect to z ($d^2H_z/dz^2$) which makes it ideal for imaging magnetic charge [25] upon the 3DASI lattice. In the present study, we focus upon the field-driven transport of magnetic charge upon the L1 and L2 layers. The volume of the individual nanowires is sufficiently high that the 3DASI system is frozen at room temperature and thus thermal energies are negligible when compared to the energy required to switch a wire.

It is insightful to first study the simplest scenarios where each sub-lattice is saturated. Optical magnetometry (see Fig S2) indicates 30 mT is well above the saturating field for each sub-lattice. Figure 2a presents an MFM image, taken at remanence following a H = 30 mT in-plane magnetic field, first applied along unit vector (1,-1,0) and subsequently along unit vector (-1,-1,0). Masks are placed over void regions to guide the eye to signal originating from L1 and L2. Unmasked data is provided in the supplementary information. Every L1-L2 vertex within the array is seen to have identical contrast. A magnified example of the contrast associated with an individual L1-L2 vertex is also shown in Fig 2b. With our choice of tip magnetisation, the bright yellow lobes indicate a positive phase associated with the stray field at magnetisation tail whilst bright red lobes indicate a negative phase associated with the stray field at magnetisation head. Focusing first upon the L1 nanowires, one can see lobes of strong positive contrast at the upper left of the nanowires and negative contrast in the lower right of the nanowires. Now focusing upon L2, strong positive contrast is seen in top right of nanowires, with negative contrast seen in bottom left. Overall, the vertex configuration is consistent with a type 2 ice-rule configuration produced by the applied field protocol. We note that near the bottom left of the L2 nanowires, faint positive contrast is seen (Labelled A). A previous investigation, which took images in reversed tip configurations identified this as an artifact [24], due to the abrupt upwards change in topography experienced by the tip at this point. Since the signal originating from the artifact is approximately a factor of two smaller than the signal originating from magnetic contrast, its presence does not impede analysis of the magnetic configuration.

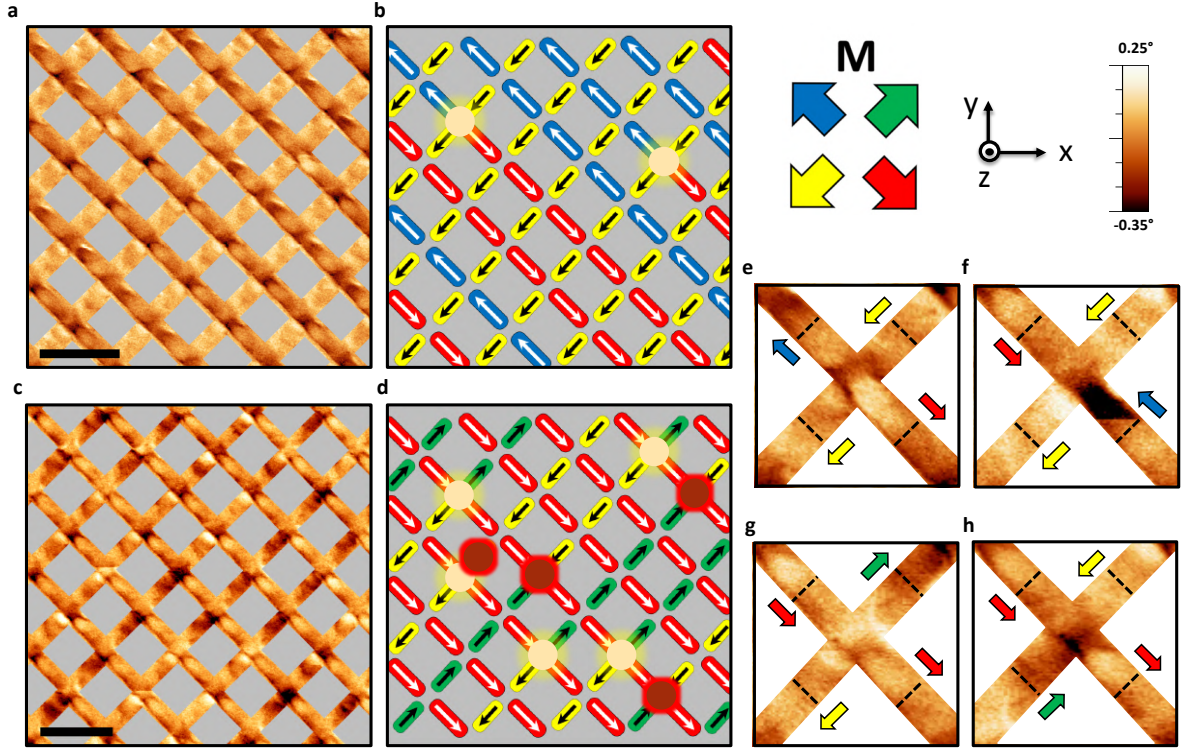

**Figure 3: Identification of monopole-excitations. (a)** MFM image taken at remanence following a saturating field along the unit vector (1,-1,0) and subsequent 9.5mT field along the unit vector (-1,1,0). **(b)** Associated vector map illustrating the magnetic configuration, monopole-excitations are annotated with bright yellow (Q = -2q) and red (Q = +2q) lobes. Each island represents a bipod, coloured with the local in-plane magnetization, as determined by key. **(c)** MFM image taken at remanence following a saturating field along unit vector (-1,-1,0) and subsequent 8.0mT field applied along (1,1,0). (d) Associated vector map illustrating the magnetic configuration, and presence of monopole-excitations **(e-h)** Magnified examples of the MFM constrast associated with L1-L2 junctions where Q = ±2q.

To demonstrate that each sub-lattice can reverse independently, we now take images after saturating fields along different principal axes. Fig 2c shows the large scale MFM image taken at remanence after a saturating field along unit vector (-1,1,0). It is clear that contrast upon L1 wires have inverted. Further inspection of the magnified example (Fig 2d) clearly shows the lobes of contrast upon L1 have indeed inverted showing the magnetization here has switched. The contrast upon L2 is found to be unchanged, as expected. The system was then returned to the initial state (Fig 2a) before a saturating field was applied along the unit vector (1,1,0). Examination of Fig 2e, now shows contrast upon every L2 nanowire has changed. Close inspection of Fig 2f now shows stronger positive contrast in bottom left and strong negative contrast in top right, suggesting the wires have switched. Overall, these results provide confirmation that L1-L2 vertices corresponding to saturated states can be identified. Our previous work [24] suggests that faint contrast is also expected at the top of L1 coordination two vertices (black dashed line in Figs 2 b,d,f) and at mid points upon L2, close to the L2-L3 junction (blue dashed line in Figs 2 b,d,f). Such contrast is expected even for uniformly magnetised states, due to a change in sign of $M_Z$ at the vertex. Upon L1, the effect of this is to smear out the edge contrast, such that fainter contrast of lower magnitude is seen at the L1 coordination two vertex. At the L2-L3 vertex, faint contrast is also seen, but we note that this is not sufficient to determine the magnetic state of the L3 layer.

With this fundamental understanding we next sought to image the magnetic configuration of vertex states observed during the switching process to determine if monopole-excitations can be identified and tracked. Figure 3a shows an MFM image following a saturating field along the unit vector (1,-1,0) and subsequent 9.5mT field along the unit vector (-1,1,0). Optical magnetometry in Fig S2 indicates this is within the field range that switching is expected upon L1. A vector map of the magnetic configuration (Fig 3b) has been produced through observations of the MFM contrast associated with each L1-L2 vertex as well as the surrounding wires. We note that there are multiple independent means to confirm the presence of a monopole. Firstly, contrast near the L1-L2 vertex is an excellent indication. If three of the four wires have contrast of the same sign, this is a monopole state. This can be further confirmed by then checking contrast upon the opposite ends of the wires. Finally, since the magnetic charge upon the wire ends closest to L1-L2, smears over the vertex area, the absolute magnitude of the phase is increased, when compared to an ice-rule state. We have used all three criteria simultaneously to identify monopoles at the L1-L2 vertex. Interestingly, so long as a well defined field protocol is used, it is also possible to infer the presence of monopoles at the L2-L3 vertex. Here, so long as L3 has been saturated, we expect this sub-lattice to be uniformly magnetised. However, if the extremities of two adjacent L2 nanowires both have positive or negative contrast, a monopole is implied at the L2-L3 vertex.

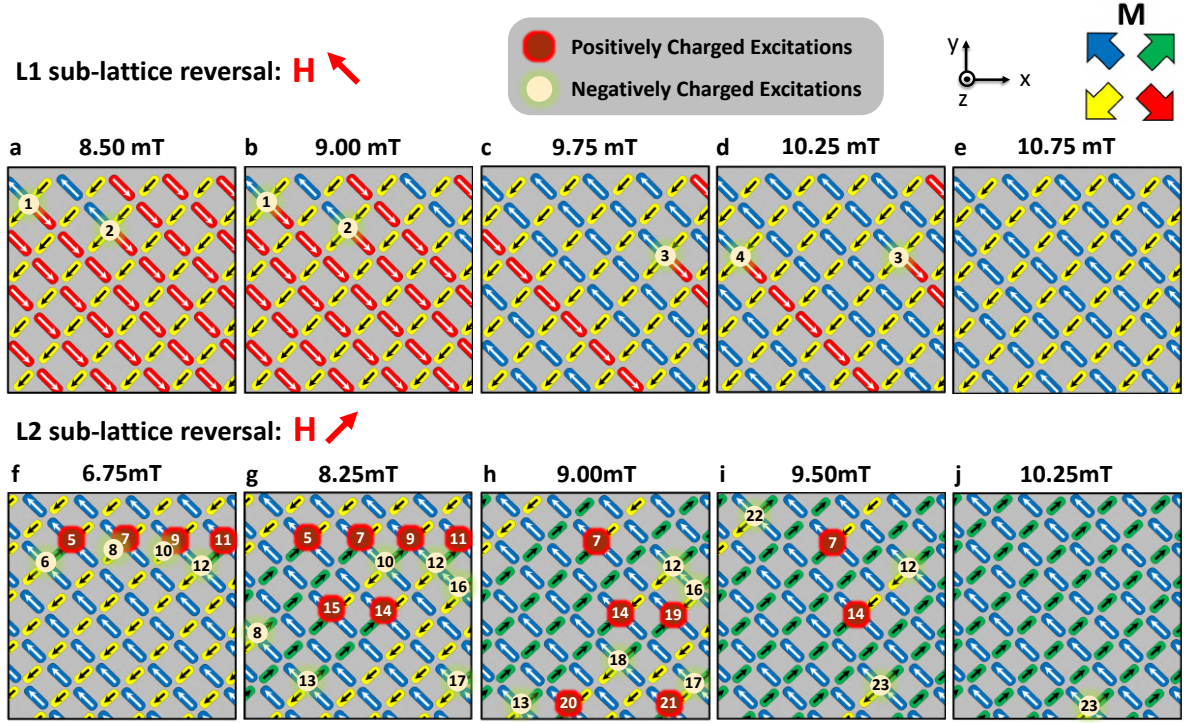

**Figure 4**: **Direct imaging of magnetic charge upon a 3D artificial spin ice system. (a-e)** Vector maps illustrating the magnetisation configuration and associated monopole excitations in five snapshots during a reversal sequence upon the L1 sub-lattice. Here a saturating field was first applied along the unit vector (1,-1,0) after which a field of 8mT was applied along the unit vector (-1,1,0). Successive images were then captured at remanence following 0.25 mT increments. Each island represents a bipod, coloured with the local in-plane magnetization, as determined by key. Each monopole excitation is assigned a unique index to track propagation between images. **(f-j)** Vector maps illustrating an equivalent reversal of the L2 sub-lattice. Here the samples was first saturated along the unit vector (-1,-1,0) after which a field of 6.50mT was applied along (1,1,0). Successive images were then captured at 0.25mT increments. Full datasets, including raw MFM images can be found in the supplementary information.

In Fig 3a-b every L1-L2 vertex in the observed area resembles one of the patterns seen in Fig 2b and d, with two exceptions. These are two monopole-excitations, each with a charge of $Q = -2q$, readily identified due to the enhanced MFM signal, which is a factor of 2 greater than the corresponding ice-rule state. Furthermore, the signal associated with the L1 wires either side of the monopoles is clearly seen to oppose, whereas the L2 wires are identical and so must be aligned. Figure 3c-d shows a similar intermediate state following a saturating field applied along unit vector (-1,-1,0) and subsequent 8.0mT field applied along (1,1,0). This allows intermediate states to be probed upon the L2 layer. Here, 9 monopoles are identified through observations of the contrast associated with each L1-L2 junction, as well as the surrounding wires. Figure 3e-h shows magnified examples of monopole-excitations with $Q = \pm 2q$, in each case one pair of colinear wires exhibits opposing contrast with respect to one another, whilst the other pair of colinear wires show matching patterns of contrast. We note that for both intermediate states (Fig 3a,c), the sub-lattice that extends along the field direction is effectively demagnetised ($M < 0.1 M_S$), so it is intriguing that a vast difference in the density of monopole-excitations is seen between the two images.

**Tracking monopole propagation on the lattice surface and sub-surface**

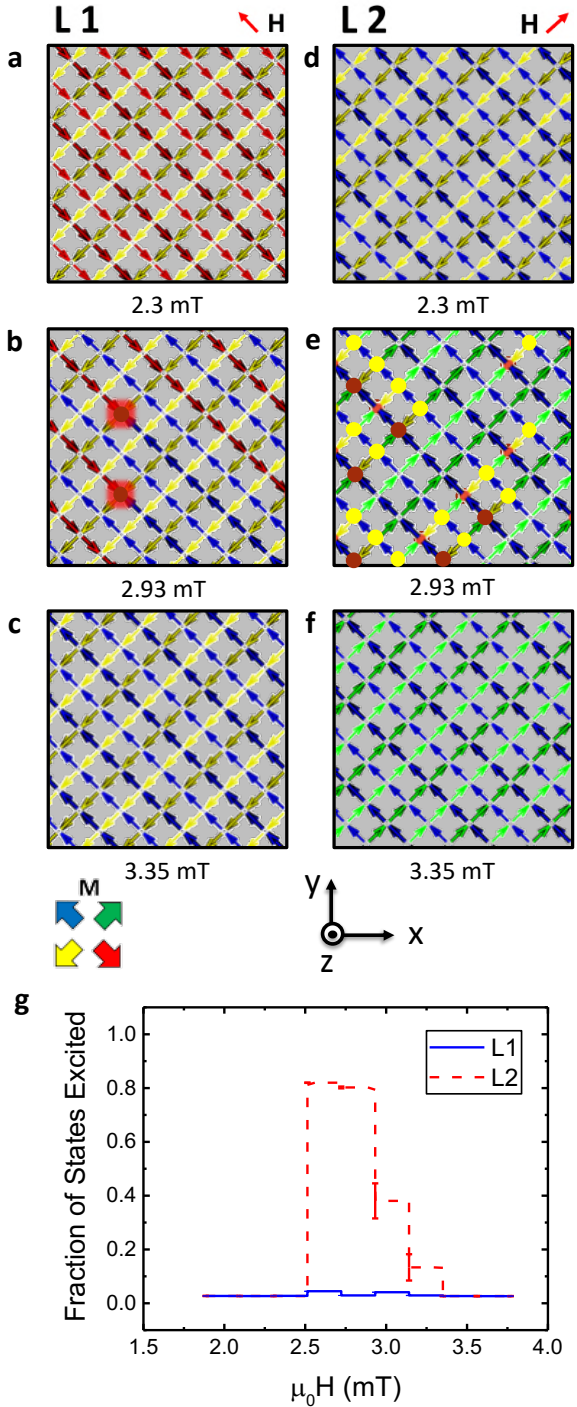

**Figure 5: Simulating the monopole dynamics upon a 3D artificial spin-ice.** (a-c) Arrow maps showing magnetisation configuration with field applied along unit vector (-1,1,0). Arrows on L1 and L2 have black borders, arrows on L3 and L4 are borderless. (d-f) Arrows maps for field applied along (1,1,0) resulting in highly correlated monopole pairs. (g) Fraction of excited states during the L1 (blue) and L2 (red) reversal sequences.

To form a more complete understanding of the monopole behavior on the surface and sub-surface, we now measure the detailed switching between two saturated states, taking images at 0.25mT intervals. To do this we carry out direct observations of the reversal sequences for the L1 and L2 sub-lattices. Figure 4 shows vector maps representing snapshots (full MFM data can be found in the supplementary information) of the switching process for the upper two layers of the lattice. Here each island corresponds to a bipod on the lattice, as defined in Fig 1d. Each image contains approximately 70 wires on L1 and 70 wires on L2, only counting those where the majority of the wire is within the measured area. Analysis herein considers wires within this $8 \times 8\ \mu m^2$ measured region, this is due to a compromise between size of the observed area and data acquisition time. Figure 4a illustrates the array after application of 8.5 mT along the unit vector (-1,1,0). This field magnitude yields the first evidence of switching along this direction. Though much of the array remains saturated, six wires (three bipods) have switched yielding two monopole states, each with charge -2q (Monopoles 1,2). In both cases, the monopoles are found at the intersection between L1 and L2. Further field increments yield additional chains of wires switching (Fig. 4b-d), with a further two negative monopoles (Monopoles 3, 4) residing at the L1-L2 junction, after which L1 reaches saturation within the sampled area (Fig. 4e).

Fig 4f illustrates the measured region after the array had been saturated along the unit vector (-1,-1,0) and a field of 6.75 mT applied in (1,1,0). Eight monopoles can be immediately seen (Monopoles 5-12), all of which seem to have appeared in pairs of ±2q. Here, five monopoles reside upon the L1-L2 junctions, whilst the remaining two reside upon L2-L3 junctions. Additional field increments lead to the creation of further monopoles (Monopoles 12-18), whilst others move along the L2 nanowires or propagate out of the measured area (Fig 4g-4j).

The differences in monopole formation upon the L1 and L2 sub-lattices is striking. Application of an external field with component along L1 yields few uncorrelated magnetic charges (Fig S10a) within the measured region, which seem to only be observed within a narrow field window (8 mT - 10.5 mT). We note that whilst this yields a net charge locally in the measured area, charge neutrality is expected across the full lattice. Analysis of the switching also shows a distinct absence of magnetic charges upon surface vertices with coordination two. On the contrary, the L2 switching leads to nucleation of many correlated pairs yielding almost equal numbers of positive and negative magnetic charges (Fig S10b), meaning the net charge within the measured area is close to zero throughout the field range (Fig S10c). The magnetic charges are also formed at a lower field (6.5mT) for the L2 sub-lattice and remain for a wider field range (6.5mT – 10.75mT).

**Modelling the 3DASI system**

Calculating the total energy density of every possible vertex state, within a micromagnetic framework (Fig S11) is an insightful exercise and provides some initial understanding of the system. Here it can be seen that the energy density to create a magnetic charge upon a coordination two, surface vertex is 3.2 times higher than that of a monopole at a coordination four vertex suggesting surface charges will be very unfavourable. To understand the significance of this within the context of switching the entire array, we carry out Monte-Carlo (MC) simulations based upon a compass needle model (see methods). This is carried out for varying surface energetic factors ($\alpha$) and quenched disorder arising from fabrication ($d_i$, see methods). A disorder of $d_i$=30% showed good agreement with switching field distributions in experimental data. The surface energetics factor ($\alpha$) essentially scales the energy required to produce a monopole upon the coordination two vertex, when compared to a coordination four vertex. A series of simulations with varying $\alpha$ are shown in Fig S14. Simulations which considered degenerate monopole surface energetics ($\alpha$=1, Fig S14) with a field applied along projection of L1, (-1,1,0) showed the presence of magnetic charges upon surface coordination two vertices and also short Dirac strings, in contrast to experimental data. Increasing the surface energetics factor to the value calculated in finite element simulations ($\alpha$=3.2), now reduces the number of magnetic charges seen upon surface coordination two vertices but Dirac string lengths are still shorter than seen in experimental results.

Fig 5a-c shows the results of MC simulations performed with enhanced surface energetics ($\alpha$=6.4) for field applied along the unit vector (-1,1,0). Upon the threshold of switching (Fig 5b), chains of islands switch upon the L1 sub-lattice producing uncorrelated monopoles and long Dirac strings as seen in the experimental data, before the majority of the array becomes

saturated (Fig 5c). Critically, charges upon surface coordination two vertices are now very rare, which is in agreement with experiment. Fig 5d-f shows MC simulations for the field aligned along unit vector (1,1,0). Here, a low field immediately produces large numbers of correlated monopole-antimonopole pairs (Fig 5e), separated by a single lattice spacing, closely aligned with the experimental data. Fig 5g summarises the simulation results by showing the fraction of excited states obtained upon L1 and L2, showing excellent qualitative agreement with the experimental data, presented in Fig S10.

## Discussion

As in all ferromagnetic materials, the 3DASI studied here passes through a field-driven state whereby the component along the field is effectively demagnetised. It is interesting to identify two main ways that this can be achieved in this novel 3D nanostructured system. The first possibility is that of local demagnetization upon each vertex, whereby the production of monopole/anti-monopole pairs locally yield a net magnetization of zero upon the relevant sub-lattice. A second possibility is the production of stripes of alternating magnetization direction, yielding complete demagnetization upon a given sub-lattice. Here magnetic charges can only be found at the stripe ends. A key quantity which will be important in determining the means of demagnetization is that of the monopole effective chemical potential, which quantifies the extent to which monopoles remain closely correlated. This is defined as $\mu^* = \mu/u$, where $\mu$ is the chemical potential of a monopole and $u = \mu_0 Q^2/4\pi a$. We note that when this value approaches half the Madelung constant (For diamond lattice, M/2=0.819) [26], a highly correlated monopole crystal is energetically favorable and hence is a possible state during the field driven dynamics. Within a simple dipolar model, we calculate (See methods) $\mu^* \approx 1.179$ for four-way junctions upon L2. Surface energetics restrict magnetic charges upon coordination two vertices, so we must consider both the high-energy, coordination two intermediate state, modulated by the factor $\alpha$, and the final state in which some energy has been spent separating the monopoles from this intermediate state (Fig. S12). The intermediate state, if stable, would imply an effective chemical potential of $\mu^*_{int} = 4.331$. Though the system must clear this energy barrier to transition to a more favorable state, it is more conventional to only consider the chemical potential with respect to the final state. Due to the less favorable Coulomb interaction of the monopoles, the effective chemical potential to produce a monopole across an L1 coordination two vertex is $\mu^* \approx 1.5661$, overall yielding a larger fraction of uncorrelated charges. Though this latter value alone may not completely justify the striking differences in monopole dynamics upon the lattice, the high energy intermediate state is far more compelling and we note that this local energy barrier (See Fig S12,S13) must be cleared to produce monopoles upon the L1 sub-lattice.

A key question that remains is the magnitude of surface energetic factor ($\alpha$) and why such large values are required in MC simulations ($\alpha$=6.4) when compared to the magnitude implied by micro-magnetics. The surface energetics in these systems arise due to a difference in how the magnetic charge is distributed for two-way and four-way junctions [24]. In both cases this will be dictated by a balance between exchange and dipolar energies. For coordination two vertices, the reduced effective dimensionality and resulting confinement produces an unfavourably large energy for monopoles upon the vertex. In contrast, the coordination four system allows the magnetic charge to spread across the vertex area, overall reducing the energy and yielding a stable monopole configuration. It is important to note that even when $\alpha$=3.2 (value indicated by MM simulations) the resulting MC simulations still bear a far closer resemblance to experiments than when enhanced surface energetics are not considered ($\alpha$=1), in terms of string length, monopole density, and density of charges upon surface coordination two vertices. However, increasing $\alpha$ beyond the value predicted by MM simulations yields an even closer resemblance to experiments, due to fundamental differences in the two methods. In particular, the MC simulations use a compass needle model, where the magnetic charge associated with each wire is distributed evenly across each needle, effectively reducing the energy barrier for surface charges to form. Therefore, a greater value of $\alpha$ is required to suppress surface charges and hence approximate the experimental observations.

In conclusion, we have demonstrated the fabrication of a 3DASI system, where the magnetic configuration upon the upper two nanowire layers can be determined. We find a striking difference in the field-driven magnetic monopole transport along two principle axes. With a field applied along the projection of surface termination, magnetic imaging shows a low number of uncorrelated monopoles during the switching, which are always found at coordination four vertices. Applying a field along the projection of L2 yields large numbers of correlated monopoles. Micromagnetic and Monte Carlo simulations, supported by simple calculations within a dipolar framework, suggest it is the difference in effective chemical potential, as well as the energy landscape experienced during surface monopole dynamics, which accounts for the measured differences. We anticipate that our study will inspire a new generation in artificial spin-ice study whereby the ground state in these 3DASI systems are explored as a function of key parameters such as magnetic moment and lattice spacing. Ultimately, this may also yield the realisation of monopole crystals as predicted in bulk spin-ice [26] or bespoke spin-ice ground states only possible in artificial systems of novel 3D geometry. By utilizing a full suite of magnetic imaging techniques including MFM, nanoscale ballistic sensing [27] and novel synchrotron-based methods [28], it is hoped that full 3D characterization of the bulk and surface will soon be possible.

**Methods**

**Fabrication**

Diamond-bond lattice structures were fabricated upon glass coverslips via two-photon lithography (TPL). Substrates were first cleaned in acetone, followed by isopropyl alcohol (IPA), and dried with a compressed air. Next, droplets of Immersol 518 F immersion oil and IPL-780 photoresist are applied to the lower and upper substrate surfaces respectively. Using a Nanoscribe Photonic Professional GT system, a polymer scaffold in the diamond-bond lattice geometry was defined within the negative tone photoresist, of dimensions 50 μm × 50 μm × 10 μm. Samples were developed in propylene glycol monomethyl ether acetate for 20 minutes, then 2 minutes in IPA, to remove any unexposed photoresist. Once again, the samples were dried with a compressed air gun.

Using a thermal evaporator, a uniform 50 nm film of Permalloy ($Ni_{81}Fe_{19}$) was deposited on the samples from above, yielding a magnetic nanowire lattice upon the polymer scaffold. This deposition requires a 0.06 g ribbon of $Ni_{81}Fe_{19}$, washed in IPA, and evaporated in an alumina coated molybdenum boat. A base pressure of $10^{-6}$ mBar is achieved prior to evaporation, the deposition rate is 0.2 nm/s, as measured by a crystal quartz monitor.

**Scanning electron microscopy**

Imaging was performed using a Hitachi SU8230 SEM with an accelerating voltage of 10kV. Images were taken from top view as well as at a 45° tilt with respect to the substrate plane.

**Magnetic force microscopy**

MFM measurements were performed in tapping mode using a Bruker Dimension 3100 Atomic Force Microscope. Commercial low moment MFM tips were magnetised along the tip axis with a 0.5 T permanent magnet. Once mounted, uniform magnetic fields could be applied parallel to each sub-lattice using a bespoke quadrupole electromagnet, which was fixed upon the surface of the AFM stage. During the application of a field, the MFM tip was positioned several mm above the scanning height, such that the tip magnetisation was not influenced. MFM data was taken at a lift height of 100 nm. Prior to capturing MFM images, feedback settings were carefully optimised to ensure sample topography was being accurately measured on the three uppermost lattice layers (L1, L2, L3).

In order to probe the transport of magnetic charge upon the 3DASI surface, the system was placed into a well-defined state by saturating the array along a principal direction (Hsat obtained via optical magnetometry, see Fig. S2). MFM images were then obtained after successive field increments in the reverse direction. MFM measures the stray field gradient $d^2H_z/dz^2$ due to magnetic charges and hence is an ideal methodology to visualise such transport across the surface [25].

**Finite element simulations**

Micro-magnetic simulations of bipod and tetrapod structures were carried out with NMAG[28], using finite element method discretisation. Geometries possessing wires with a crescent-shaped cross-section were designed such that the arcs subtend a 160° angle. The inner arc is defined from a circle with 80 nm radius corresponding to the 160 nm lateral feature size of the TPL system. The outer arc is based on an ellipse with an 80 nm minor radius and 130 nm major radius, yielding a thickness gradient with a peak of 50 nm. The length of all wires is set to 780 nm, due to computational restraints. All geometries were meshed using adaptive mesh spacing with a lower limit of 3 nm and upper limit of 5 nm. Simulations numerically integrated the Landau-Liftshitz equation upon a finite element mesh. Typical $Ni_{81}Fe_{19}$ parameters were used, i.e. $M_S = 0.86 \times 10^6$ $Am^{-1}$, $A = 13 \times 10^{-12}$ $Jm^{-1}$ with zero magnetocrystalline anisotropy.

**Monte Carlo simulations**

Each nanomagnet is modelled as an infinitesimally thin compass needle with a uniform, linear magnetic moment density *mL*, with exceptions to this being made at coordination number two vertices. The moment orients along the long axis of the island. The interaction between compass needles are equivalent to two equal and opposite magnetic charges with charges *m/L* placed at their ends with exceptions for the coordination two vertex energy. The coordination four energy calculated by the micromagnetic simulations corresponds to compass needles with $L = 0.92a$ while the micromagnetic energies for coordination two imply an interaction strength 3.2 times stronger than those between other charges. The compass needle energy obeys Coulomb's law with corrections for experimental considerations:

$$E_{ij} = \alpha_{ij} d_i \frac{\mu_0 m^2}{4\pi L^2} \left[ \frac{1}{|\boldsymbol{r}_{ai} - \boldsymbol{r}_{aj}|} - \frac{1}{|\boldsymbol{r}_{ai} - \boldsymbol{r}_{bj}|} - \frac{1}{|\boldsymbol{r}_{bi} - \boldsymbol{r}_{aj}|} + \frac{1}{|\boldsymbol{r}_{bi} - \boldsymbol{r}_{bj}|} \right] - d_i \boldsymbol{m}_i \cdot \boldsymbol{B}$$

Where $\boldsymbol{r}_{ai}$ and $\boldsymbol{r}_{bi}$ are the locations of the positive and negative magnetic charge on the ith nanomagnet, $\boldsymbol{B}$ is the external field applied either in the L1 or L2 direction with simulated magnitudes ranging between 1.89 and 3.77 mT, $\boldsymbol{m}_i$ is each nanomagnet's magnetic moment with amplitude m = MV (with *M* being the saturation magnetization, and V the nanomagnet volume), $\mu_0$ is the magnetic permeability, and $L$ = 1000 nm is the island length. The magnetization was chosen to be $M = 850$ kA m$^{-1}$ in agreement with previous studies on nanoislands fabricated in the same manner. $\alpha_{ij}$ is a factor which increases the interaction strength between *ij* pairs at coordination number 2 vertices on the L1 sub-lattice with respect to *ij* pairs at all other vertices, this captures the enhanced surface energetics indicated by micro magnetic simulations. Reversals were simulated with $\alpha_{ij}$ = 1, 3.23, and 6.45 at coordination number 2 vertices on L1, $\alpha_{ij} = 6.45$ was found to yield the closest agreement to experimental observations, all other vertices were consistently defined with $\alpha_{ij} = 1$. Site disorder $d_i$ is drawn from the distribution $P(d) = \frac{1}{\sigma\sqrt{2\pi}} e^{-\frac{(d-1)^2}{2\sigma^2}}$ ($\sigma = 30\%$ is found to yield good agreement with experimental data in this case). This disorder arises due to subtle variations in 3D nanowire geometry. Systems of the same dimensions as the experiment, one unit cell deep and 15 by 15 unit cells wide and long, were simulated with 20 replicas apiece.

The simulations began in the saturated state seen in our experiments. A random spin was selected and the energy to flip that spin was calculated. The corresponding spin flip was carried out if the energy lost exceeded a threshold energy corresponding to the magnetic coercivity, an algorithm equivalent to the Metropolis method with zero temperature and used in prior spin ice studies[17,29]. Flips were attempted 10 times the number of total spins, sufficient for equilibration. The resulting arrow maps were plotted in Fig. 5a-f and Fig. S13. Additionally, the number of excited states was recorded from these final arrow maps and plotted as a function of field in Fig. 5g.

**Dipolar approximation calculations**

One can define an effective chemical potential $\mu^* = \mu/u$, where $u = \mu_0 Q^2/4\pi a$. Here $\mu$ is calculated within the dipolar approximation. Magnetic moments are located upon a diamond-bond lattice. The energy of interaction between moments can then be approximated as:

$$E_{12} = u \frac{|\hat{m}_1 \cdot \hat{m}_2 - 3(\hat{m}_1 \cdot \hat{r})(\hat{m}_2 \cdot \hat{r})|}{\left|\frac{r}{a}\right|^3}$$

This choice of deunitization places the physical value of the lattice parameter $a$ in $u$. All geometric factors are then derived from a lattice where the lattice constant is one. Considering any two spins within a coordination four geometry yields $E_{12} = \frac{5}{6\sqrt{2}} u$. For an ice-rule state there are four low energy pairs and two high energy pairs yielding $E_{ice} = -\frac{5}{3\sqrt{2}}u$. A doubly charged monopole can be created by flipping a single spin. Each monopole state has three low energy pairs and three high energy pairs, making $E_{monopole} = 0$. The energy landscape for creation of an L2 monopole is shown in Fig S12a. The chemical potential can thus be obtained via $\mu = (E_{monopole} - E_{ice}) = \frac{5}{3\sqrt{2}}u$. Therefore, the effective chemical potential, $\mu^*$, for a coordination four geometry is $\frac{5}{3\sqrt{2}}$ or about 1.179. In units of Kelvin this yields ~700K, as expected for a system in the frozen regime.

Now turning to monopole nucleation upon the L1 sub-lattice, the overall energy landscape is shown in Fig S12 (Top), with a more detailed depiction shown in Fig S13. Surface energetics restrict the presence of magnetic charges on the coordination two vertices. Imposing this constraint now requires the resulting monopole-antimonopole pair to separate now $\frac{2\sqrt{2}}{\sqrt{3}}$ lattice spacings apart. The energy of monopole creation is the same because two "monopole" coordination four vertices are converted to a higher energy state, but the long-range interactions increase the energy. Approximating this via the dumbbell model's long-range interactions, $E_{lr} = \frac{u}{r_{charge}/a}$ where $r_{charge}$ is the distance between the monopoles, the energy required to produce a surface monopole rises by the difference between the interaction of monopoles a lattice spacing apart and monopoles separated further. This raises the cost by $u\frac{2\sqrt{2}-\sqrt{3}}{2\sqrt{2}}$, yielding an effective chemical potential, $\mu^*$, of $\frac{5}{3\sqrt{2}} + \frac{2\sqrt{2}-\sqrt{3}}{2\sqrt{2}}$, or approximately 1.566. From a dynamic perspective, this state must be created by either monopoles traversing the bulk with multiple spin flips or generating a monopole on the system's surface. The former is prohibited dynamically because it would only be possible with a mix of L1 and L2 direction spin flips. This is forbidden because the external field responsible for the flips is only applied in one direction at a time. The latter option raises the energy further as the coordination two monopole is created with both different geometry and an increased surface energetic factor when compared to the coordination four monopole. The emergence of a coordination two monopole is from a single favorable interaction, $-\frac{5}{6\sqrt{2}}\alpha u$, to a single unfavorable interaction, $\frac{5}{6\sqrt{2}}\alpha u$, each of which are modified by the α factor described in the Monte Carlo simulations. This requires an energy change of $\frac{5}{3\sqrt{2}}\alpha u$. Averaging this with the coordination four monopole gives the average, reduced cost per monopole in this intermediate state, $\mu_{int}^* = \frac{5}{3\sqrt{2}}\frac{\alpha+1}{2}$. The system effectively leaps from a reduced, average energy of $-\frac{5}{6\sqrt{2}}\frac{\alpha+2}{2}$ to an average of $\frac{5}{12\sqrt{2}}\alpha$. For our experimentally confirmed $\alpha = 6.4$, this implies a transition state costing $\mu_{int}^* = 4.331$.

**Magneto-optical Kerr Effect Magnetometry**

A 0.5 mW, 637 nm wavelength laser was expanded to a diameter of 1 cm, passed through a Glan–Taylor polarizer to obtain an s-polarized beam, then focused onto the sample using an achromatic doublet (f = 10 cm), to obtain a spot size of approximately 10 μm². During the source-to-sample path the laser is attenuated, approximately reducing the power by a factor of 4. The reflected beam was also collected using an achromatic doublet (f = 10 cm) and passed through a second Glan–Taylor polarizer, from which the transmitted signal was directed onto an amplified Si photodetector, yielding the Kerr signal. After magneto-optical Kerr effect data was captured from the nanowire lattice, a second dataset was obtained from the substrate film. The film data is scaled to the lattice data and subtracted off. This corrected for any Kerr signal contributions originating from the film, during the lattice measurements.

**Funding:** SL acknowledges funding from the Engineering and Physical Sciences Research Council (Grant number: EP/R009147/1/)

**Author contributions:** SL conceived of the study and wrote the first draft of the manuscript. Sample fabrication and Magnetic force microscopy was carried out by AM. Monte Carlo simulations and chemical potential calculations were carried out by MS. Finite element simulations were performed by AM and AV. Optical magnetometry was carried out by JA and MH. All authors contributed to writing the final manuscript.

**Competing interests:** The authors declare no competing interests.

**Data and materials availability:** Information on the data presented here, including how to access them, can be found in the Cardiff University data catalogue.